# Jitter Characterization of a Dual-Readout SNSPD

D. F. Santavicca, B. Noble, C. Kilgore, G. A. Wurtz, M. Colangelo, D. Zhu, and K. K. Berggren, *Fellow, IEEE*

*Abstract*—To better understand the origins of the timing resolution, also known as jitter, of superconducting nanowire single-photon detectors (SNSPDs), we have performed timing characterizations of a niobium nitride SNSPD with a dual-ended readout. By simultaneously measuring both readout pulses along with an optical timing reference signal, we are able to quantify each independent contribution to the total measured jitter. In particular, we are able to determine values for the jitter due to the stochastic nature of hotspot formation and the jitter due to the variation of the photon detection location along the length of the nanowire. We compare the results of this analysis for measurements at temperatures of 1.5 K and 4.5 K.

*Index Terms*—nanowire single-photon detectors, jitter, superconducting photodetectors, superconducting device noise

## I. Introduction

THE superconducting nanowire single-photon detector (SNSPD) is capable of detecting single visible and near-IR photons with near unity detection efficiency, ~ns reset times, and ~10 ps timing resolution. As a result, these detectors are attractive for some of the most demanding single-photon applications [1]. The excellent timing resolution is critical for many such applications. For example, accurate determination of the arrival time of a photon limits the security of quantum key distribution [2] and the data rate of quantum photonic circuits that rely on single-photon sources and the coherence properties of entangled photons [3]. It also sets the resolution in time-of-flight laser ranging (LIDAR) [4]. As a result, understanding the limits of timing resolution in SNSPDs and engineering devices with improved timing resolution has become an active area of research [5]-[11].

A typical SNSPD consists of a compact nanowire meander with one end connected to ground and the other end connected to the input of a 50 Ω microwave amplifier. In recent work, it was shown that connecting a readout circuit to each end of the nanowire can be used to determine the location of the photon detection along the length of the nanowire based on the time difference between the two output pulses [12],[13]. It was also shown that averaging the time of the two output pulses can increase the timing resolution, i.e. decrease the jitter [6]. This effect was most pronounced in longer nanowires, as the improved jitter results from minimizing the timing uncertainty originating from the distribution of photon detection locations along the length of the nanowire. In the present work, we show that this dual-readout approach can be extended to enable a full characterization of all the independent contributions to the total device jitter.

## II. Jitter Contributions

The jitter of each detector output can be found experimentally by measuring the time difference between each output pulse and an optical timing reference signal. The jitter is defined as the full-width at half-maximum (FWHM) of a Gaussian fit to a histogram of many such time difference measurements. This total jitter has contributions from (a) the jitter from the noise of the readout amplifier $J_{amp}$, (b) the jitter due to the distribution of photon detection locations along the length of the nanowire $J_{geo}$, (c) the jitter due to the stochastic nature of the hotspot formation $J_{hotspot}$, (d) the jitter due to the noise on the timing reference signal $J_{timing}$, and (e) the jitter due to the finite optical pulse width $J_{opt}$. The contributions from (a), (d), and (e) arise from the measurement system while the contributions from (b) and (c) arise from the SNSPD device itself. These sources of jitter are all assumed to be uncorrelated and hence their contributions add in quadrature. If we call $J_1$ the total jitter on output 1 and $J_2$ the total jitter on output 2, then we have:

$$J_1^2 = J_{amp1}^2 + J_{geo}^2 + J_{hotspot}^2 + J_{timing}^2 + J_{opt}^2 \quad (1)$$
$$J_2^2 = J_{amp2}^2 + J_{geo}^2 + J_{hotspot}^2 + J_{timing}^2 + J_{opt}^2 \quad (2)$$

For simplicity, we assume that these sources of jitter are all Gaussian random variables, although this may not strictly be the case.

If we consider a photon absorbed in the nanowire at some location $x$, where the nanowire extends from $x = 0$ to $x = L$, then the output pulse reaches the end at $x = 0$ after a time $t_1 = x/v$ and the output pulse reaches the other end after a time $t_2 = (L-x)/v$, where $v$ is the pulse velocity on the nanowire [6]. If we average $t_1$ and $t_2$, we get $t_{avg} = (t_1+t_2)/2 = L/(2v)$, which does not depend on $x$. Hence the fluctuations in $t_{avg}$, which correspond to a jitter $J_{avg}$, do not depend on $J_{geo}$. Averaging the two output pulses also reduces the jitter due to the noise of each amplifier by a factor of 2, and hence $J_{avg}$ is given by

$$J_{avg}^2 = J_{amp1}^2/4 + J_{amp2}^2/4 + J_{hotspot}^2 + J_{timing}^2 + J_{opt}^2 \quad . \quad (3)$$

We can also measure the time difference between output 1 and output 2. We call the FWHM of this distribution the differential jitter $J_{diff}$. This does not depend on $J_{timing}$. It will also not have any contribution from $J_{opt}$ or $J_{hotspot}$, as the photon arrival time and the hotspot formation for each detection event are common to both outputs. Taking the difference of $t_2$ and $t_1$,

This work is supported by NSF grants ECCS-1509253 and ECCS-1509486. D. Z. is supported by the National Science Scholarship from A*Star Singapore. *(Corresponding author: Daniel F. Santavicca.)*

D. F. Santavicca, B. Noble, C. Kilgore, and G. A. Wurtz are with the Department of Physics, University of North Florida, Jacksonville, FL 32224 (e-mail: daniel.santaviccca@unf.edu).

M. Colangelo, D. Zhu, and K. K. Berggren are with the Research Laboratory of Electronics, Massachusetts Institute of Technology, Cambridge, MA 02139.

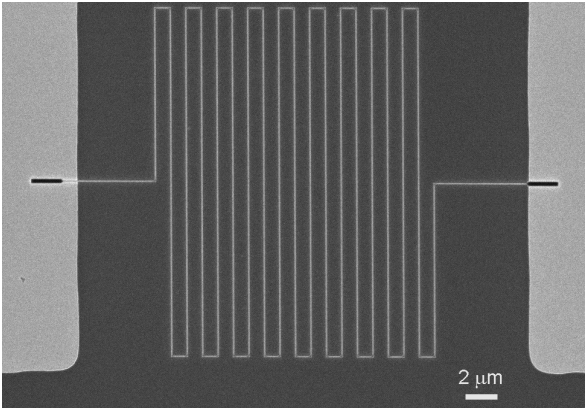

Fig. 1. Scanning electron micrograph of SNSPD device studied in this work. The NbN nanowire meander connects to gold leads on either side. The gold leads are the center conductor of a 50 Ω CPW transmission line.

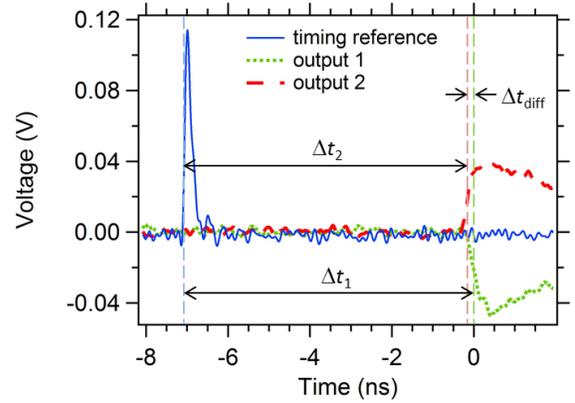

Fig. 2. Example of a recorded detection event at $T$ = 1.5 K. The detector produces two output voltage pulses with opposite polarity. An optical timing reference signal is also recorded.

we get $t_{\text{diff}} = t_2 - t_1 = (L-2x)/v$. Hence fluctuations in the photon detection location $x$ will produce twice the fluctuation in $t_{\text{diff}}$ as they do in $t_1$ and $t_2$. As a result, we have for $J_{\text{diff}}$

$$J_{\text{diff}}^2 = 4J_{\text{geo}}^2 + J_{\text{amp1}}^2 + J_{\text{amp2}}^2. \quad (4)$$

$J_1$, $J_2$, $J_{\text{avg}}$, and $J_{\text{diff}}$ can be determined directly from timing measurements. $J_{\text{opt}}$ can be determined from the optical pulse width. $J_{\text{amp1}}$, $J_{\text{amp2}}$, and $J_{\text{timing}}$ can be determined by measuring the output noise voltage on output 1, output 2, and the timing signal, respectively, and then dividing by the pulse slew rate, i.e. the slope of the pulse at the voltage threshold where the arrival time is defined. Once these quantities have been determined, one can solve for the two remaining quantities, $J_{\text{hotspot}}$ and $J_{\text{geo}}$, using (3) and (4).

## III. DEVICE DETAILS AND EXPERIMENTAL SETUP

The device fabrication process has been described previously [14]. Briefly, an ≈ 6 nm thick niobium nitride (NbN) film was sputtered onto a high-resistivity silicon substrate with a native oxide. Electron beam lithography and a $CF_4$ etch were used to pattern a nanowire with a width of 90 nm and a total length of 438.6 μm, seen in Fig. 1. The nanowire was patterned into a meander with 1 μm pitch, and the corners were rounded to minimize current crowding. Each end was connected to the center conductor of a gold 50 Ω coplanar waveguide. The nanowire has $T_C$ = 8.44 K and a sheet resistance of 453 Ω/sq at 14 K. The measured critical current is 14.7 μA at 4.5 K and 21.3 μA at 1.5 K. The critical current density indicates that this is not a significantly constricted device. This is confirmed by a measurement of the kinetic inductance as a function of bias current, which found that, at 1.5 K, the inductance increases by a factor of 1.28 compared to its zero-current value just below the switching current [14],[15].

The device is mounted in the same microwave sample holder used in [14]. The sample holder is placed inside the vacuum can of a helium cryostat with a base temperature of 1.5 K. Both sides of the device connect to room temperature through 50 Ω semi-rigid coaxial cables. At room temperature, each end is connected to a bias tee and a low-noise 50 Ω amplifier with a 0.01-3.0 GHz bandwidth, 34 dB of gain, and a 1.4 dB noise figure. The amplifier outputs are coupled to a 6 GHz, 20 GSa/s real-time oscilloscope. The dc bias line is heavily filtered to prevent low frequency noise from coupling to the device, which can lead to a reduced switching current.

The device was illuminated using a sub-ps pulsed 800 nm laser. The pulse width at the laser output was approximately 20 fs. The optical signal is coupled to a Thorlabs 780HP single-mode optical fiber. A 90/10 fiber splitter is used to couple the larger power to a 6 GHz InGaAs photodetector for use as a timing reference and the smaller power to the device inside the cryostat. The total fiber length between the fiber input and the device is approximately 2 m. The fiber dispersion at 800 nm is specified as -117 ps/nm/km, resulting in a pulse width of approximately 10 ps at the device. In future work, this pulse width could be decreased through the use of a bandpass filter at the fiber input. The end of the optical fiber is positioned approximately 1.5 cm above the device, resulting in significant geometrical attenuation.

Detection events were recorded at 1.5 K and 4.5 K. A bias current of 18 μA was used at 1.5 K and 11 μA was used at 4.5 K. In each case, the count rate was measured as a function of bias current to ensure that the device was being operated with saturated internal detection efficiency. We also verified that the average number of detection events per laser pulse was much less than one to ensure that we are in the single-photon regime.

## IV. RESULTS AND ANALYSIS

At each temperature, a total of $10^4$ detection events were recorded. For each event, we recorded both SNSPD output pulses as well as the timing reference signal, as seen in Fig. 2. The average of each of these outputs is used to determine the maximum slope and the corresponding voltage. This is used as the voltage threshold at which the pulse arrival times are defined. At 4.5 K, the maximum slope was $7.077 \times 10^7$ V/s for output 1, $7.996 \times 10^7$ V/s for output 2, and $1.058 \times 10^9$ V/s for the timing signal. At 1.5 K, the maximum slope was 1.993 × $10^8$ V/s for output 1, $2.094 \times 10^8$ V/s for output 2, and $1.185 \times 10^9$ V/s for the timing signal.

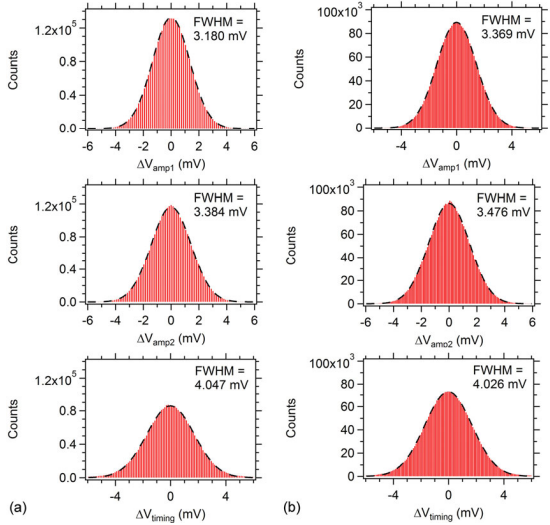

Fig. 3. Histograms of the measured voltage noise, from top to bottom, of output 1, output 2, and the timing signal at (a) 4.5 K and (b) 1.5 K. Each histogram is fit to a Gaussian function to determine the FWHM.

The voltage noise, defined as the difference between the instantaneous voltage and the average voltage, of each detector output and the timing signal was found for each detection event using the data away from each peak. A histogram was made for each set of voltage noise data, and the histogram was fit to a Gaussian function to determine the FWHM. These data, shown in Fig. 3, show excellent agreement with the Gaussian fits. The FWHM was then divided by the maximum slope values to determine the corresponding jitter; these values for $J_{amp1}$, $J_{amp2}$, and $J_{timing}$ are given in Table 1.

The histograms for the timing differences $\Delta t_1$, $\Delta t_2$, and $\Delta t_{diff}$, as illustrated in Fig. 2, along with the time difference between the average pulse and the timing reference, $\Delta t_{avg}$, are shown in Fig. 4. At 4.5 K, the data were fit to a Gaussian. At 1.5 K, the histograms exhibit a noticeable asymmetry, and so they were fit to an exponentially modified Gaussian function [8]. In both cases, the jitter was taken as the FWHM of the Gaussian function.

With $J_{amp1}$, $J_{amp2}$, $J_{diff}$, $J_{avg}$, and $J_{timing}$ determined experimentally, and $J_{opt}$ estimated from the optical pulse width, we then solve for $J_{geo}$ and $J_{hotspot}$ using (3) and (4). These values are listed in Table 1. As a consistency check, we can use the calculated values of $J_{geo}$ and $J_{hotspot}$ in (1) and (2) to calculate $J_1$ and $J_2$, which can then be compared to the measured values. These calculated values are given in parenthesis next to the measured values in Table 1.

In order to better understand $J_{geo}$, we performed a microwave-frequency measurement of the first-order $\lambda/2$ nanowire self-resonance using the same technique as [14]. As the wavelength $\lambda$ is twice the nanowire length $L$ at the first-order self-resonance frequency $f_0$, we can solve for the phase velocity $v = f_0\lambda = 2f_0L$. At 4.5 K and a bias current of 11 μA, the first-order resonance was at 9.11 GHz, which corresponds to $v = 7.99 \times 10^6$ m/s or $0.0266c$, where $c$ is the speed of light in free space. At 1.5 K and a bias current of 18 μA, the first-order resonance was at 9.08 GHz, corresponding to $v = 7.96 \times 10^6$ m/s or $0.0265c$. This velocity suggests a significantly larger value of $J_{geo}$ than determined from our timing data, as 438.6 μm / $0.0265c$ = 55 ps. We can understand this discrepancy if we consider the effect of pulse dispersion on the nanowire [16]. To this end, we simulated pulses generated by resistive hotspots at various locations along the nanowire using the AX-IEM solver in AWR Design Environment. In this simulation, the nanowire sheet impedance was set to the value determined from matching a simulation of the nanowire first-order self-resonance to the measured data [14]. The hotspot was assumed to have a resistance of 3.5 kΩ and a duration of 0.4 ns. We simulated the output pulses for hotspots at 28 locations at regular intervals along the length of the nanowire. The results are shown in Fig. 5. In the inset, we show the corresponding FWHM jitter at different threshold voltages. We see that this gives values considerably less than 55 ps and close to the experimental values found for $J_{geo}$.

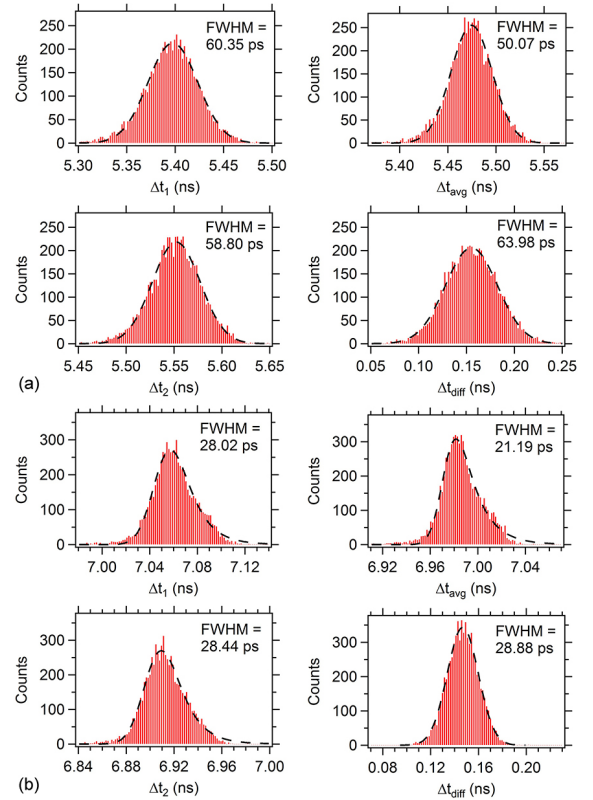

Fig. 4. Histograms of the measured values of $\Delta t_1$, $\Delta t_2$, $\Delta t_{avg}$, and $\Delta t_{diff}$ from $10^4$ individual measurements at (a) 4.5 K and (b) 1.5 K. Data at 4.5 K are fit to a Gaussian, while data at 1.5 K are fit to an exponentially modified Gaussian. In both cases, the jitter is determined from the FWHM of the Gaussian.

TABLE I
RESULTS OF JITTER CHARACTERIZATION

|  | $T = 4.5$ K | $T = 1.5$ K |
|---|---|---|
| $J_1$ | 60.4 ps (60.4 ps) | 28.0 ps (25.7 ps) |
| $J_2$ | 58.8 ps (58.5 ps) | 28.4 ps (25.5 ps) |
| $J_{avg}$ | 50.1 ps | 21.2 ps |
| $J_{diff}$ | 64.0 ps | 28.9 ps |
| $J_{amp1}$ | 44.9 ps | 16.9 ps |
| $J_{amp2}$ | 42.3 ps | 16.6 ps |
| $J_{timing}$ | 3.8 ps | 3.4 ps |
| $J_{opt}$ | 10 ps | 10 ps |
| $J_{geo}$ | 8.4 ps | 8.3 ps |
| $J_{hotspot}$ | 37.9 ps | 14.0 ps |

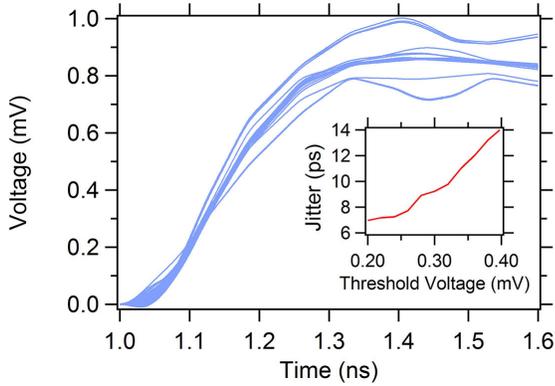

Fig. 5. Simulated output pulses for 28 hotspot locations at regular intervals along the nanowire. Inset: Jitter calculated from these pulses as a function of threshold voltage.

## V. CONCLUSION

The calculated values of $J_1$ and $J_2$ agree well with the measured values, particularly at 4.5 K. The slightly less good agreement at 1.5 K may come from additional uncertainty in the exponentially modified Gaussian fits. At 4.5 K, we see a significantly larger overall jitter, which is related to a larger amplifier jitter as well as a larger hotspot jitter. The change in the amplifier jitter is due to the lower bias current used at 4.5 K, which results in a smaller pulse amplitude and hence a smaller slope.

$J_{geo}$ is very similar at the two temperatures, which is consistent with our determination of a similar phase velocity on the nanowire for the two temperatures and bias currents. The value of $J_{geo}$ is approximately consistent with a simulation of pulses produced by different hotspot locations along the nanowire. This simulation shows the significant role that pulse dispersion, which arises from the meandered nanowire geometry [16], plays in the value of $J_{geo}$.

The larger hotspot jitter at 4.5 K is presumably related to larger thermal fluctuations leading to greater variation in the hotspot formation process. Recent theoretical work has sought to quantify the effects of Fano fluctuations, variation in the photon absorption location across the width of the nanowire, and nanowire spatial inhomogeneity on the jitter [9],[10]. None of these mechanisms seem to be able to explain the large temperature dependence seen in $J_{hotspot}$. Future models may benefit from the use of a non-deterministic Monte Carlo approach.

The value of $J_{hotspot}$ at 1.5 K is larger than the total jitter recently reported in a 5 μm long NbN SPNSD at a similar temperature and photon wavelength [8]. This suggests that the nanowire geometry plays a role in the hotspot jitter. Indeed, recent work found that the presence of bends in the nanowire leads to increased jitter [7]. Ultimately, the goal is to understand the lower limit of jitter for practical devices with high optical efficiency. Such understanding will likely benefit from further studies of the dependence of the hotspot and geometrical jitter on device parameters such as the nanowire geometry and material. We hope that the approach described in this paper for quantifying the different contributions to the total device jitter using a dual-readout detector will facilitate such studies.

## ACKNOWLEDGMENT

The authors would like to thank C. Bunker and Q.-Y. Zhao for helpful discussions.